\newcommand{\diracslash}[1]{#1\llap{/\kern2pt}}

\newcommand{\be}{\begin{equation}}
\newcommand{\ee}{\end{equation}}
\newcommand{\bea}{\begin{eqnarray}}\index{\footnote{}}
\newcommand{\eea}{\end{eqnarray}}
\newcommand{\ba}[1]{\begin{array}{#1}}
\newcommand{\ea}{\end{array}}

\newcommand{\Rmnum}[1]{\expandafter\@slowromancap\Romannumeral #1@}
\documentclass[doublespacing,preprint]{revtex4-1} 
\usepackage{hyperref} 
\usepackage{amsmath}
\usepackage{graphicx}
\usepackage{epsfig}
\usepackage{amsmath}
\usepackage{textcomp}

\begin{document}
\title{Effects of a static electric field on two-color photoassociation between different atoms}
\author{Debashree Chakraborty$^1$ and Bimalendu Deb$^{1,2}$}
\address{$^1$Department of Materials Science, Indian Association for the Cultivation of Science,
 Jadavpur, Kolkata 700032, INDIA}
\address{$^2$Raman Center for Atomic, Molecular and Optical Sciences, Indian Association for the Cultivation of Science,
Jadavpur, Kolkata 700032, INDIA}
\begin{abstract}
We study non-perturbative effects of a static electric field on two-color photoassociation of different atoms.
A static electric field induces anisotropy in scattering between two different atoms and hybridizes field-free rotational states of heteronuclear dimers
or polar molecules. 
In a previous paper [D. Chakraborty $\it {et.}$ $\it {al.}$, J. Phys. B 44, 095201 (2011)], 
the effects of a static electric field on one-color photoassociation between different atoms has been described through 
field-modified ground-state scattering states, neglecting electric field effects on heteronuclear diatomic bound states. 
To study the effects of a static electric field on heteronuclear bound states, 
and the resulting influence on Raman-type two-color photoassociation between different atoms in the presence of a static electric field,  
we develop a non-perturbative numerical method to calculate static electric field-dressed heteronuclear bound states.
We show that the static electric field induced scattering anisotropy as well as hybridization of rotational states strongly influence two-color photoassociation spectra, leading to significant enhancement in PA rate and large shift.
In particular, for static electric field strengths of a few hundred kV/cm, two-color PA rate involving high-lying bound states in electronic ground-state increases by several orders of magnitude even in the weak photoassociative coupling regime. 
\end{abstract}

\maketitle
\section{Introduction}
Continuum and bound states between different atoms can be readily manipulated with a static electric field. 
Because, a static electric field can interact with the permanent dipole moment of the collision complex or the diatomic molecule formed from two different atoms.
It has been shown that due to their permanent electric dipole moment,
the response of heteronuclear molecules to an external electric field is
stronger than the response of homonuclear molecules \cite{krems1,krems2}.
With the recent advancement in the production of cold polar molecules \cite{bigelow,marcassa,k.k.ni,kerman,sage,kraft,deiglmayr,wedimullerchemrev2012,takekoshipra2012}, manipulation of continuum of scattering states as well as bound states between cold atoms with both static and dynamic electromagnetic fields is of prime importance.
In particular, the effects of a static electric field on photoassociation (PA) of cold different atoms are important for controlling atom-molecule transitions in polar systems.    
Various methods of coherent control of PA have been theoretically discussed in a recent review by Koch and Shapiro \cite{shapirochemrev2012}.

In a previous paper \cite{debashree}, we have demonstrated large enhancement of single-photon PA rate due to a static electric field-induced anisotropic scattering resonances of a different atom-pair. 
An electric field couples a large number of partial waves of relative motion of the atom-pair. This highly anisotropic interaction between the atoms leads to large modification of quantum scattering states at low energy.
At certain specific electric field strengths, low energy scattering resonances appear in collision between two heteronuclear ground-state atoms. These are multichannel resonances where channels are degenerate and refer to the partial waves of relative motion between the atoms. 
For Li + Cs collision, the first resonance occurs at electric field $\cal E$ = 1298 kV/cm. Contributions from at least 10 coupled partial waves to the total scattering cross section need to be included in order to produce the first resonance.   
Near this anisotropic resonance, one-color PA spectral profile shows large enhancement in spectral intensity with significant modifications in rotational structure \cite{debashree}. 
PA rates can also be enhanced by controlling shape resonances with strong nonresonant light as suggested by Gonzalez-Ferez and
Koch \cite{Gonzalezpra2012}.
An intense laser field-induced modification of interaction between identical ground-state atoms basically arise from polarizability due to induced dipole moment, while the interaction between different atoms can be modified due to coupling of laser field with both permanent and induced dipole moments.

The purpose of this paper is to study the effects of a strong static electric field on bound states of a polar molecule and how these effects are manifested in 
PA in the presence of static electric field. In the previous work \cite{debashree}, we have investigated the effects of a static electric field only on the continuum states of a pair of two ground-state different atoms and the consequent modifications of one-color PA, disregarding electric field effects on excited molecular bound states. 
However, an electric field of the order of a few hundred kilovolt is strong enough to affect nonperturbatively not only the continuum states but also the bound states of a polar molecule both in electronically ground and excited states. 
Therefore, in treating Raman-type two-color photoassociation of different atoms in the presence of a static electric field, it is essential to take into account the effects of the field on rotational and vibrational states.
Previously, investigations into the effects of an external field on rotational and vibrational states have been carried out by several authors \cite{meyenn,2004,2005,gonzalezpra2007,gonzaleznjp2009,gonzalezpccp2011} in various contexts.
Gonzalez-Ferez and Schmelcher \cite{2004} have developed a hybrid computational technique based on discrete variable representation (DVR) and basis set expansion to calculate full ro-vibrational spectra of diatomic heteronuclear molecules in strong electric fields \cite{gonzaleznjp2009}. This computational method has been applied to study the effects of a static electric field on the formation of heteronuclear molecules by one-photon stimulated emission from collisional states of different atoms \cite{gonzalezpra2007}.

Here we develop an alternative method to account for the nonperturbative effects of an external static electric field on rotation-vibration coupled bound states in electronically excited or ground polar molecule. 
To calculate bound states of the coupled-channel Schr$\ddot{o}$dinger equation of a polar system in the presence of a static electric field, we have used renormalized Numerov method \cite {johnsonjcp1978}.  
Due to the hybridization of the rotational motion, the
binding energies are found to alter significantly with electric field strength. 
Our main aim here is to investigate the effects of such electric field-induced hybridization on two-color PA. 
We find that for static electric field strengths on the order of a few hundred kV/cm, both one- and two-color PA rates increase by several orders of magnitude even in the weak photoassociative coupling regime.
This occurs particularly for high-lying vibrational levels in excited molecular potential. Static electric field induced effects also lead to large shifts in PA spectra. 
We further show that, even with weak PA lasers, saturation can be achieved by tuning electric field near anisotropic resonances. This occurs due to large enhancement of free-bound Franck-Condon (FC) overlap near anisotropic resonances. 
Since free-bound stimulated linewidth is proportional to the square of FC integral, a large enhancement of this integral due to anisotropic resonances leads to saturation. 
Due to this saturation effect, a splitting occurs in one-color PA spectra. In case of two-color PA, when the bound-bound transition is tuned near one of the peaks of one-color PA at saturation, a further splitting occurs resulting in a three peak PA spectrum.
Our results show that, in the absence of electric field, one- and two-color PA probability is negligibly small in comparison to that in the presence of electric field induced anisotropic resonances.
We choose Li + Cs heteronuclear system for illustrating our results. 
This system has large permanent dipole moment compared to other systems of current interests. 
The paper is presented in the following way. 
In section \MakeUppercase{\romannumeral 2}, we develop a method of calculation of electric field-dressed bound states of two different atoms. We apply this method to calculate bound state of LiCs dimer in the presence of a static electric field. 
One- and two-color PA in the presence of a static electric field are described in section \MakeUppercase{\romannumeral 3}.
The results are discussed in section \MakeUppercase{\romannumeral 4}. The paper is concluded in section \MakeUppercase{\romannumeral 5}.

\section{Electric field-dressed bound and continuum states}
In this section we describe electric field-dressed bound and continuum states of different atoms. 
In the Born-Oppenheimer approximation, on averaging over the electronic motion, the total Hamiltonian reduces  to  
\bea 
\hat{H} = - \frac{\hbar^2}{2 \mu } \nabla_{{\mathbf R}}^2 + \hat{H}_{\rm{hf}} + \hat{V}_{c}(R) + \hat{V}_{\cal E}
\eea 
where $\mu$ is the reduced mass and 
$\nabla_{{\mathbf R}}$ is the Laplacian operator that describes relative motion between  the two nuclei of the atoms with $\mathbf{R}$ being the relative 
position vector, 
$\hat{H}_{\rm{hf}}$ represents hyperfine interaction of the two separated atoms,  $\hat{V}_{c}$ refers to the central interaction consisting of 
singlet and triplet adiabatic potentials and $\hat{V}_{\cal E} = - \vec {\cal E} . \vec{D}(R)$ is the interaction of the permanent dipole moment 
$\vec{D}$  with the applied static electric field  $\vec {\cal E}$. Both $\hat{V}_{c}$ and $\vec{D}$ are functions of internuclear separation $R$. 
As mentioned earlier, the interaction $\hat{V}_{\cal E}$ between the static 
electric field and the permanent dipole moment of a polar system such as heteronuclear dimer or colliding atom-pair is essentially 
anisotropic.
As a result, when the bound or continuum wave function of such a dimer or the colliding atom-pair interacting with a static electric field is expanded 
in the rotational basis, $\hat{V}_{\cal E}$ can be expressed in a matrix form. In our previous work \cite{debashree}, continuum wave function of the relative motion between 
two heteronuclear ground-state atoms was expanded in 
terms of partial waves represented by $\ell$  which is the rotational quantum number of the field-free relative motion between the 
two free atoms  with $m_{\ell}$ being its projection on the space-fixed $z$-axis which is chosen to be along the direction of electric field. Since an electric field defines an axis of rotational symmetry on which the projection of the total angular momentum of the molecule is constant, m$_{\ell}$ remains a good quantum number, however $\hat{V}_{\cal E}$ couples different $\ell$ with $\Delta \ell = \pm 1$. 
Since a strong electric field can nonperturbatively affect 
the bound states of polar dimers in both excited and ground electronic configurations that can be accessed by one- or two-color PA, it is then essential 
to account for the change in bound states due to the electric field.

Here we develop a method for calculation of electric field-dressed bound states of a diatomic polar molecule. As in the case of continuum states, 
the different rotational angular momenta of field-free bound states will be coupled due to $\hat{V}_{\cal E}$. We denote rotational quantum numbers of the 
field-free bound states by $J$ and $M$ which are the counterparts of $\ell$ and $m_{\ell}$ of the continuum state. The hyperfine interaction mixes 
the singlet and triplet potentials leading to hyperfine channel potentials. For simplicity we consider only one 
hyperfine channel. The effective Hamiltonian in the rotational basis in the presence of a static electric field can then be expressed as 
\bea
 H = -\frac{\hbar^2}{2\mu R^2}\frac{\partial}{\partial R}(R^2\frac{\partial}{\partial R}) + \frac{J^2(\theta,\phi)}{2\mu R^2} + V_c(R)- \vec {\cal E} . \vec D(R) \eea
where the angles $\theta$ and $\phi$ specify the orientations 
of the interatomic axis in the space-fixed coordinate frame. 
The single-channel potential is represented by $V_c(R)$. The dipole moment function $D(R)$ can be represented as
\bea
D(R)= \sum_S\sum_{M_S}\mid S M_S\rangle d_S(R)\langle S M_S\mid
\eea
where $d_S$(R) denotes the dipole moment function in the different spin states $S$ which takes the value 0 and 1 for singlet and triplet state, respectively.
Assuming that the electric field is applied along the $z$-axis,
electric field couples different $J$ with difference $\pm$1 while its projection on $z$-axis $M$ remains a good quantum number.

Thus the problem 
of finding a proper bound state of polar dimer in the presence of a static electric field concerns essentially to evaluate 
a multicomponent or multichannel bound state wave function where components or channels refer to the field-free rotational states. 
In case of solving for anisotropic scattering wave function, we earlier used renormalized multichannel 
Numerov-Cooley algorithm. Though here also we follow essentially the same procedure as used for calculating anisotropic scattering state,
there are difficulties in case of calculating multichannel bound states, stemming from the fact that 
all the components of the bound state should satisfy boundary conditions at both $R=0$ and $R \rightarrow \infty$. In case of 
multichannel scattering problem, the Numerov-Cooley code can be propagated in one way from $R \simeq 0$ to large $R$ and then 
the numerically calculated solution matrix can be matched with multichannel scattering boundary conditions at asymptotically large 
$R$ to evaluate scattering matrices and multichannel wave function. In case of multichannel bound state 
calculation the situation is quite different as we show in \cite{supplemental}.

Under single hyperfine-channel approximation, using rotational basis, time-independent Schr$\ddot{o}$dinger equation of a polar dimer interacting 
with the static electric field  can be written in the matrix form  
\bea
[ {\bf I}\frac{d^2}{dR^2} + {\bf Q}(R)]{\bf \Psi}(R)=0
\eea
where ${\bf I}$ is the unit matrix and ${\bf Q}(R)=(\frac{2\mu}{\hbar^2})[E{\bf I}-{\bf V}(R)]$.
Here ${\bf V}(R)$ is the symmetric potential matrix with its matrix element
\bea 
V_{JJ'} = \left [ V_c + \frac{J(J+1)}{R^2} \right ] \delta_{JJ'} + \langle J\mid \vec D. \vec{\cal E}\mid J'\rangle.
\eea 
The wave function ${\bf \Psi}(R)$ is a square matrix whose columns are linearly independent. 
In the separated atom limit ($R\rightarrow\infty$) the Eq. (4) reduces to an equation in matrix form for multichannel scattering and $J$ should be replaced by partial-waves $\ell$ of relative motion between the two atoms. 
Note that the threshold of atom-atom potential does not primarily depend on the applied static electric field since in the asymptotic limit spherically symmetric atoms do not possess any permanent dipole moment. It is only at short separations where two different atoms are in a state of collision or form a bound state, a permanent dipole moment emerges in the collision complex or the bound state. However, the threshold can change by the second order Stark effect due to the applied static electric field. In our calculations we do not take into account this second order Stark effect, because this can be safely neglected compared to the first order effects due to permanent dipole moment as we discuss in the section IV.

The wave function in a column form can be expressed as some linear combination of the columns of ${\bf \Psi}(R)$ such as
\bea 
{\bf \psi}(R)= {\bf \Psi}(R).{\bf C} 
\eea
where $\bf C$ is a column vector of constant coefficients.
According to Gordon procedure \cite{Gordon}, a trial value of energy is chosen and Eq. (4) is first numerically integrated inward from the outer boundary and then outward from the inner boundary 
to a common matching point $R_m$.  
This matching point is conveniently chosen near the last antinode close to outer turning point.
The wave function calculated this way will not be in general continuous at $R_m$. 
To converge to a specific eigenvalue $E_n$, a two-fold procedure is followed. The first part of this procedure consists of calculating the number of nodes in the outward solution (there is no node in the inward integration) to isolate a single eigenvalue. 
If the number of nodes is greater than the number of nodes $n$ of the eigenfunction corresponding to the eigenvalue $E_n$, the trial energy is too high and if it is less than $n$ the trial energy is too low compared to the eigen energy E$_n$.
A bisection method \cite{johnsonjcp1978} can be used to isolate a single eigenvalue with a specified node count, within a small energy interval. 
Once it is found that the eigenvalue $E_n$ lies within some range of energy such as $E_l<E_n<E_h$; then the trial energy $E$ is set equal to
\bea
 E=\frac{E_h+E_l}{2}.  
\eea
Equation (4) is then integrated to count the number of nodes in the outward solution. 
If the node count is greater than $n$ set $E_h$=$E$ and if it is lower, set $E_l$=$E$. 
A new trial energy is calculated using equation (7) and the process is repeated until $E$ is close enough to
$E_n$ so that the number of node count is equal to $n$. When the node count is equal to $n$, we use the log derivative matrix method to converge to the specific eigenvalue. 
The log derivative matrix is defined as
\bea
    {\bf y(R)} = {\bf \Psi'}(R){\bf \Psi^{-1}}(R)
\eea  
where ${\bf \Psi}(R)$ is the solution of Eq. (4).
The matrix Numerov algorithm can be formulated by a three term recurrence relation \cite{supplemental}.

In the field-free case each state is characterized by its vibrational ($v$), rotational ($J$) and magnetic ($M$) quantum numbers.
In the presence of a static electric field, the multichannel bound wave function are characterized by $v$, $M$ and another quantum number $j$ which corresponds to the rotational
quantum number $J$ in the absence of the electric field. This means that in our method, vibrational quantum number remains a good quantum 
number, though the bound state wave function has several $R$-dependent components corresponding to different rotational coupling with 
all the components having the same number of nodes. Thus we write
\bea
{\bf \psi}_{vjM}=\sum_{J=0}^{N-1}{\bf \psi}^{{\cal E},j}_{v,J,M}(R)Y_{JM}(\hat {R})
\eea
where $Y_{JM}(\hat {R})$ is the spherical harmonics. ${\bf \psi}^{{\cal E},j}_{v,J,M}(R)$ is the $J$th component of the field dressed bound state ${\bf \psi}_{vjM}$ evolving from field-free $j$ rotational state.
It is to be noted that the sum in Eq. (9) should contain an infinite number of terms, but for all practical purposes one can truncate this series upto the first $N$ terms for which results are convergent.
For ${\cal E}$ = 0, ${\bf \psi}^{{\cal E},j}_{v,J,M}(R)$ will be equal to ${\bf \psi}_{v,J=j,M}(R)$ and all other components with 
$J \neq j$ will vanish.
This wave function satisfies the normalization condition $\int_{0}^{\infty}\sum_J|{\bf \psi}^{{\cal E},j}_{v,J,M}(R)|^2dR$=1. 

In essence, the bound state 
wave function of (9) is a superposition of all the rotational states where the superposition coefficients are the components 
of the field-dressed radial wave function with a particular vibrational number. 
For ${\cal E} \rightarrow 0$ the anisotropic field-dressed bound state is reduced to the isotropic field-free one.
There is a shift in energy eigenvalue due to field-dressing. 
Our method thus allows to calculate 
various $R$-dependent rotational components of the multichannel dressed bound state. This would be in turn useful to calculate bound-bound or free-bound 
transition dipole matrix elements that will contain the static electric field-induced effects. 
Because of superposition of different rotational states, laser induced transition amplitude will also be 
a superposition of various rotational components allowed by the electric dipole transition selection rules.

\begin{table}
\caption{The change $\Delta'_{v',j',M'}$ in GHz in binding energy of the field-dressed excited bound state (${v',j',M'}$) with respect to its field-free binding energy for different electric field strengths in kV/cm.
$\Delta_{v,j,M}$ ($\cal E$) denotes the similar change in ground bound state ($v,j,M$).}
\begin{tabular}{c  c  c  c  c  c  c  c  c  c  c  c  c  c  c  c   c  c  c  c  c   c   c    c   c   c   c   c  c  c  c  c  c  c  c  c  c  c}
\hline
& \vline& $\cal E$ &\vline& $\Delta'_{26,1,0}$ &\vline& $\Delta'_{26,2,0}$ &\vline& $\Delta'_{4,1,0}$ &\vline& $\Delta_{28,0,0}$ & \vline & $\Delta_{28,1,0}$ & \vline & $\Delta_{28,2,0}$ &\vline& $\Delta_{4,0,0}$&\vline&\\
\hline
&\vline& 50 &\vline& -0.19  &\vline&  -0.03   &\vline&   -8.55 & \vline & -11.1 & \vline & -1.29& \vline & -0.09 &\vline& -133.2 &\vline&\\
\hline
&\vline& 100 &\vline& -6.87 &\vline&  -1.36    &\vline&   -39.25 & \vline & -22.2 & \vline & -2.06& \vline & -0.84 &\vline& -62.08 &\vline&\\
\hline
&\vline& 500 &\vline& -12.46 &\vline&  -2.85    &\vline&  -5.86 &\vline & -26.27 & \vline & -4.09& \vline & -2.44 &\vline& -199 &\vline&\\
\hline
&\vline& 1000 &\vline& -30.14 &\vline&  -5.3    &\vline&  49.62 & \vline & -26.83 & \vline & -7.24& \vline & -6.17 &\vline& -67.4 &\vline&\\
\hline
&\vline& 1298 &\vline& -52.02 &\vline&  -16.47   &\vline&  3.41 & \vline & -28.78 & \vline & -12.15& \vline & -11.18 &\vline& -73.77 &\vline&\\
\hline
\end{tabular}
\end{table}
\begin{table}
\caption{Electric field variations of $\nu_{v,j,M,v',j',M'}$ is given 
along with the change $\Delta\nu$ from the corresponding field-free value for the transition between states evolving from field-free $v$=28, $j$=0 and $v'$=26, $j'$=1.}
\begin{tabular}{c  c  c  c  c  c  c  c  c  c  c  c  c  c  c  c c c  c c c c c c c c c}
\hline
&\vline& $\cal E$ &\vline& $\nu_{v,j,M,v',j',M'}$ (THz) &\vline& $\Delta\nu$ (GHz) &\vline&\\
\hline
&\vline& 0&\vline&360.469 &\vline& 0 &\vline&\\
&\vline&50 &\vline&360.480 &\vline& 11&\vline&\\
&\vline&100 &\vline& 360.488&\vline& 19&\vline&\\
&\vline&500 &\vline&360.483 &\vline&14&\vline&\\
&\vline&1000 &\vline&360.466 &\vline&-3&\vline&\\
&\vline&1298 &\vline&360.436 &\vline&-33&\vline&\\ 
\hline
\end{tabular}
\end{table}

Before concluding this section, we wish to state a few important features of the electric field-dressed scattering states which are relevant for our later discussion on PA.
Although, electric-field dressed continuum states are described previously \cite{debashree}, for completeness we reproduce some aspects of them.
The scattering wave function in the presence of a static electric field is given by \cite{debashree}
\bea
\psi_{\ell,m_{\ell}}(E,R) = \sum_{\ell',m_{\ell'}}\phi_{\ell'\ell}(E,R)Y_{\ell',m_{\ell'}}(\hat {R})\eea
where $\ell$ and $\ell'$ denote the incident and the scattered partial waves, respectively and $m_{\ell}$ and $m_{\ell'}$ are the corresponding magnetic
quantum numbers.
The partial waves $\ell, m_{\ell}$ are used to denote the angular momenta of the scattering states whereas the rotational quantum numbers $J$, $M$ refer to angular momenta of bound states. 
Unlike field-free case, the electric field-modified scattering wave function of equation (10) is essentially anisotropic. 
This has significant implications in both one- and two-color PA spectrum as described in the next two sections. 
Single-color PA spectrum at low laser intensity is proportional to the square of the overlap integral between ground-state scattering and excited-state bound wave functions. 
As mentioned earlier, the electric field-induced modification of the scattering wave function primarily occurs at short separations where PA transitions are usually dominant.
The scattering amplitude or cross section is extracted from the asymptotic behavior of the scattering wave function at large separations.
%====================================================================
%Figure 1
\begin{figure}
\includegraphics[width=4.25in]{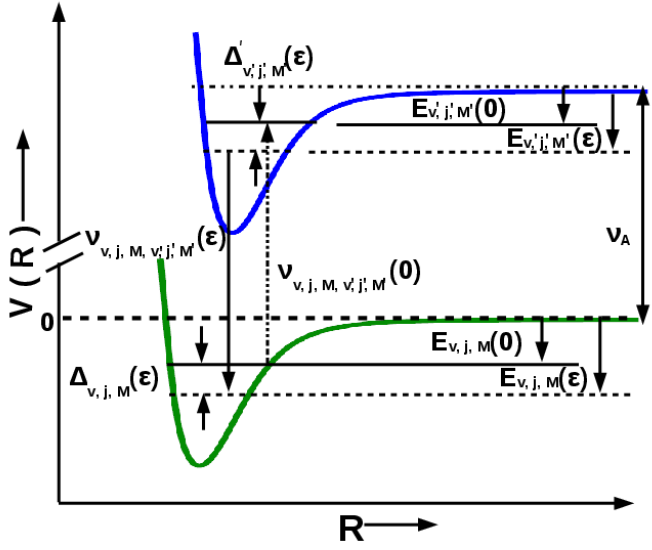}
\caption{Schematic diagram displaying the binding energies $E_{v',j',M'}$ and $E_{v,j,M}$ of the excited and the ground bound state, respectively for both field dressed ($\cal E \neq$ 0) and field-free ($\cal E$ = 0) cases. The bound-bound transition frequencies are given by $\nu_{v,j,m,v',j',M'}$($\cal E$) and $\nu_{v,j,m,v',j',M'}$(0). Also shown are the electric field-induced shifts $\Delta'_{v',j',M'}$($\cal E$) and $\Delta_{v,j,M}$($\cal E$) of the excited and the ground bound states, respectively.} 
\end{figure}
%=====================================================================
To reveal how field-modified interatomic collision cross section at different energies affect one- and two-color PA spectral profiles, we calculate total scattering cross section given by 
\bea
\sigma = 4\pi\sum_{\ell {\ell'}}\sum_{m_\ell m_{\ell'}}\mid t_{\ell m_\ell}^{\ell' m_{\ell'}}\mid ^2\eea
where $t_{\ell m_\ell}^{\ell' m_{\ell'}}$ is the T-matrix element. 
We have earlier found a one-to-one correspondence between field-induced anisotropic resonances and the enhancement in one-color PA rate \cite{debashree}. Here we examine whether such effect also occurs for two-color PA.

\section{One- and two-color photoassociation in the presence of a static electric field} 
One-color photoassociation process creates a free-bound transition pathway between the scattering state 
$\mid \psi_{\ell m_{\ell}}(E)\rangle$ of a colliding pair of ground state atoms and an excited ro-vibrational state $\mid {\bf \psi'}_{v', J', M'}\rangle$. 
Here we discuss how one- and two-color PA of two different atoms is influenced by the electric field.  
A PA spectrum is described as the rate of loss of atoms due to the excited molecular level populated by PA process. 
PA rate coefficient \cite{napolitano,julienne} is given by
\bea K_{PA}^{n} = \left\langle\frac{\pi v_{rel}}{k^2}\sum_{\ell, m_{\ell}}\mid S_{PA}^{n}(E;\ell, m_{\ell})\mid^2\right\rangle\eea 
Where $S_{PA}^{n}$ represents S-matrix element for one- ($n$ = 1) or two-color ($n$ = 2) PA transition and $\langle....\rangle$ implies an averaging over thermal velocity distribution.
%==========================================================================================================================================
%Figure-2
\begin{figure}
\includegraphics[width=4.25in]{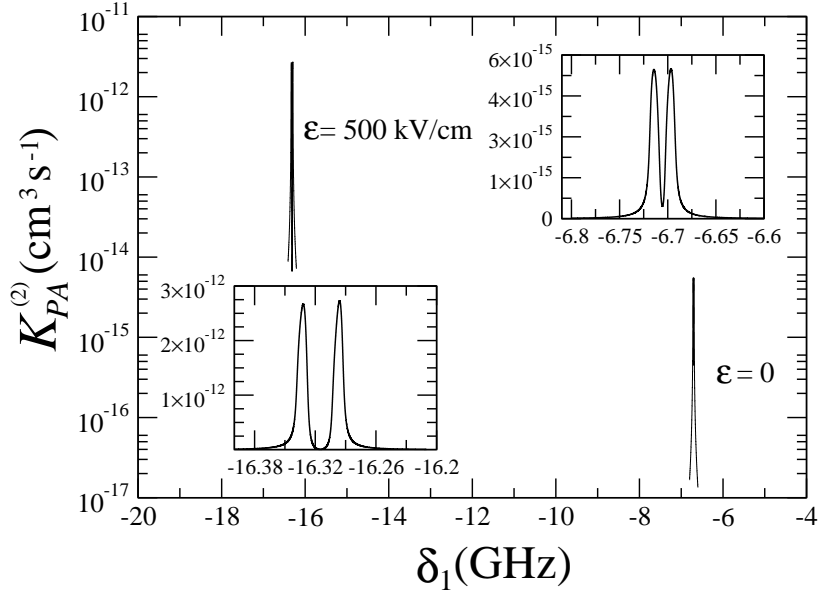}
\caption{Two-color PA rate $K_{PA}^{(2)}$ is plotted as a function of detuning $\delta_1$ (in GHz) of laser-1 at $\cal E$ = 0 and 500 kV/cm for transition between states evolving from field-free $J'$=1 of $v'$ = 26 and $J$= 0 of $v$ = 28, keeping laser-2 fixed at bound-bound resonance for $I_2$ = 1mW/cm$^2$ and $I_1$ = 1W/cm$^2$. 
The two-peak spectral shapes are clearly displayed in the two insets. 
}
\end{figure}
%==========================================================================================================================================
For sufficiently low laser intensities, the one-photon PA probability is given by
\bea \mid S_{PA}^{1}\mid^2 = \frac{\hbar^2\gamma_s\Gamma_{\ell,m_{\ell}}(E,I_1)}{[(E-\Delta_1)^2+\hbar^2((\Gamma +\gamma_s)/2)^{2}]}\eea 
where $\gamma_s$ represents the natural linewidth of the excited state.
The detuning of laser-1 is given by $\Delta_1$ = $E_{v', j', M'}$($\cal E$) - $\hbar\delta_1$ where $\delta_1$ = $\omega_{1}-\omega_A$ is the
frequency offset between laser frequency $\omega_{1}$ and atomic resonance frequency $\omega_A$ and $E_{v', j', M'}$($\cal E$) is the binding energy of the field-dressed multichannel 
excited bound state measured from the threshold $\hbar \omega_A$ of the excited potential as schematically shown in figure 1.  
The partial stimulated linewidth $\Gamma(E,\ell,I_1)$ is given by 
\bea \Gamma_{\ell,m_{\ell}}(E,I_1) = \frac{\pi I_1}{\hbar\epsilon_{0}c}\mid\langle {\bf \psi'}_{v',j',M'}(\vec {R})\mid \vec{D}_t(R).\hat{\cal E}_{L1}\mid \psi_{\ell,m_{\ell}}(E,\vec{R})\rangle\mid^2\eea
where $\vec{D}_t(R)$ is the transition dipole moment and $\hat{\cal E}_{L1}$ is the unit vector giving the electric field polarization of laser-1 with intensity $I_1$,
$\epsilon_{0}$ and c are the vaccum permitivity and speed of light, respectively.
${\bf \psi'}_{v',j',M'}(\vec{R})$ represents the field-dressed multichannel excited bound state as given in Eq. (9). The anisotropic scattering state $\mid \psi_{\ell,m_{\ell}}(E,\vec{R})\rangle$ as given in Eq. (10) is a superposition of all the scattered partial waves for the incident partial wave $(\ell,m_{\ell})$.
%=================================================================================================================================================================
%Figure-3
\begin{figure}
\includegraphics[width=4.25in]{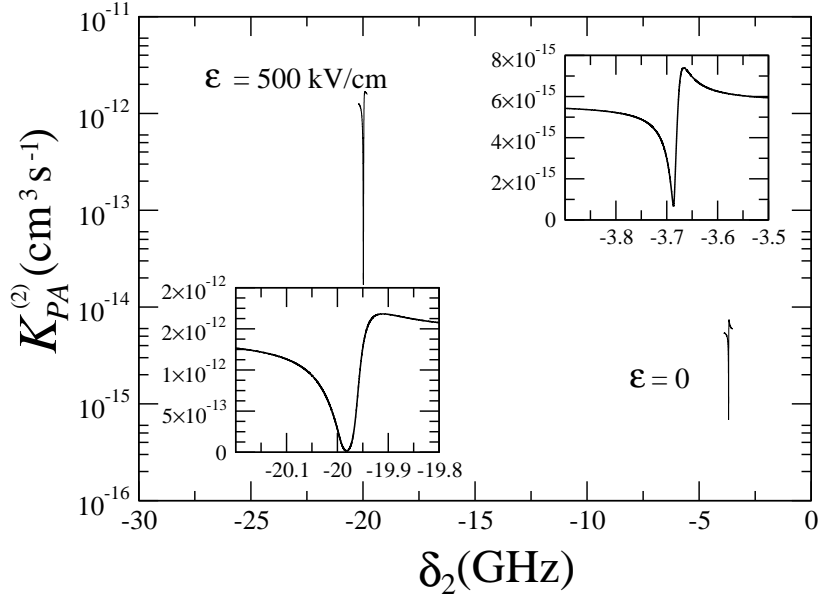}
\caption{Two-color PA rate $K_{PA}^{(2)}$ is plotted as a function of detuning $\delta_2$ (in GHz) of laser-2 at $\cal E$ = 0 and 500 kV/cm for transition between states evolving from field-free $J'$=1 of $v'$ = 26 and $J$= 0 of $v$ = 28, keeping laser-1 fixed at free-bound resonance for $I_1$ = 1W/cm$^2$ and $I_2$ = 1mW/cm$^2$. 
}
\end{figure}
%=============================================================================================================================
It is worthwhile to mention that in presence of the electric field, free-bound stimulated linewidth
$\Gamma_{\ell,m_{\ell}}(E,I_1)$ is modified significantly. 
In absence of the electric field, for a particular incident partial wave $\ell$, only one scattered partial wave contributes since both incident and scattered partial waves are same.
But in presence of a static electric field, different partial waves become coupled and thus for a particular incident partial wave there are possibilities to scatter into different outgoing partial waves. 
As a result, sum over all incident and scattered partial waves is to be
taken to evaluate PA rate. The total linewidth $\Gamma$ = $\sum_{\ell,m_{\ell}}\Gamma_{\ell,m_{\ell}}$ is a sum over partial linewidths.

In case of bound states, different rotational angular momenta become coupled in presence of the electric field and the bound-bound Rabi coupling $\Omega_{12}$ is also modified significantly. 
The two-color stimulated Raman PA probability is given by 
\bea 
\mid S_{PA}^{2}\mid^2 = \frac{\hbar^2(E-\Delta_2)^2\gamma_s\Gamma_{\ell,m_{\ell}}}{[(E-\Delta_+)(E-\Delta_-)]^2 + \hbar^2((\Gamma +\gamma_s)/2)^{2}(E-\Delta_2)^2}&.
\eea 
The second laser with intensity $I_2$ splits a single PA resonance into a pair of peaks located near the energies 
\bea
 \Delta_{\pm} = \frac{1}{2}(\Delta_1+\Delta_2)\pm\frac{1}{2}\sqrt{(\Delta_1-\Delta_2)^2 + 4\hbar^{2}\Omega_{12}^2}.
\eea 
Laser-2 is detuned by $\Delta_2 = E_{v, j, M}$($\cal E$)-$\hbar(\omega_1-\omega_2)$,
where $\omega_2$ is the frequency of the second laser.
%==========================================================================================================================================
%Figure-4
\begin{figure}
\includegraphics[width=4.25in]{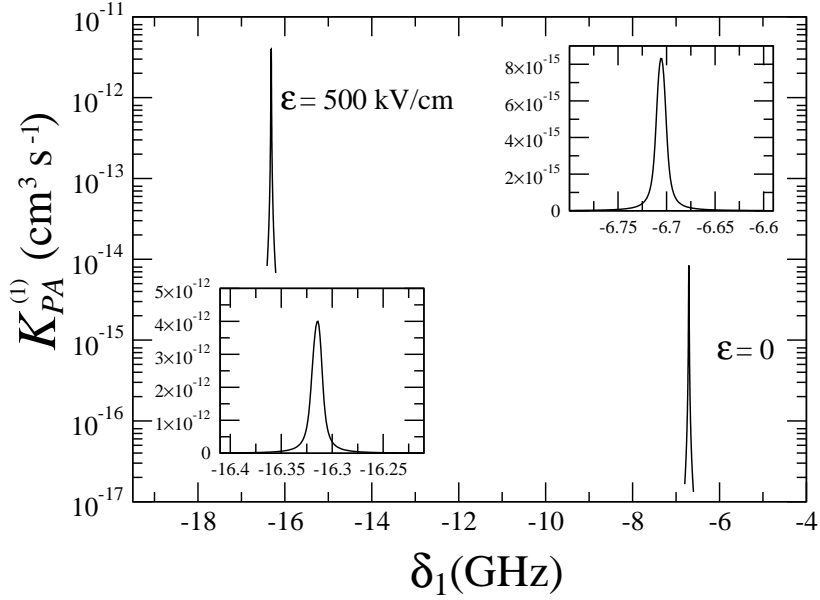}
\caption{The one-color PA rate $K_{PA}^{(1)}$ (in unit of cm$^3$s$^{-1}$) is plotted as a function of detuning $\delta_1$ (in GHz) at $\cal E$ = 0  and 500 kV/cm for transition to states evolving from field-free $J'$=1 of $v'$ = 26. The intensity $I_1$ is set at 1W/cm$^2$.
The two insets show the zoomed view of the spectral lines.}
\end{figure}
%==========================================================================================================================================
The bound-bound Rabi frequency $\Omega_{12}$ is given by
\bea\Omega_{12} = \frac{1}{\hbar}\left(\frac{I_2}{4\pi\epsilon_0c}\right)^{1/2}\mid\langle {\bf \psi'}_{v',j',M'}(\vec{R})\mid \vec{D}_t(R).\hat{\cal E}_{L2}\mid {\bf \psi}_{v,j,M}(\vec{R})\rangle\mid\eea 
where $\hat{\cal E}_{L2}$ stands for the unit vector of electric field polarization of the second laser.
The molecular Rabi coupling is proportional to the bound-bound Franck-Condon (FC) factor.
In the next section we present numerical results that reveal significant modifications of PA spectrum due to static electric field effects.

\section{Results and discussions}
As a prototype model we consider PA between Li and Cs to
form LiCs polar molecule in the presence of a static electric field. The ground state potential data for LiCs are taken from \cite{stannumpra2007}.
An analytical expression for ground-state dipole
moment function of Li + Cs collision complex is given by Li and Krems \cite{krems2}, approximating the numerical data computed by Aymar and Dulieu \cite{aymer}.  
%==============================================================================================================================================================
%Figure-5
\begin{figure}
\includegraphics[width=4.25in]{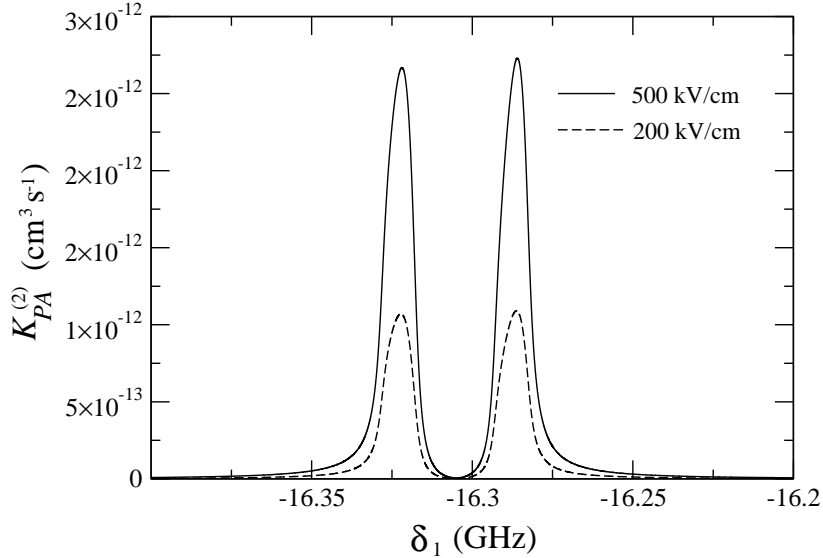}
\caption{Two-color PA rate $K_{PA}^{(2)}$ is plotted as a function of detuning $\delta_1$ (in GHz) of laser-1 at $\cal E$ = 500 kV/cm (solid line) and 200 kV/cm (dashed line)for transition between states evolving from field-free $J'$=1 of $v'$ = 26 and $J$= 0 of $v$ = 28. 
}
\end{figure}
%==============================================================================================================================================================
This is given by 
\bea
d_S(R) = D_0 \exp \left[ -\alpha(R - R_e)^2\right], \eea
with the parameters $R_e$ = 7.7 $a_0$, $\alpha$ = 0.1 $a_0^{-2}$, and $D_0$ = 6 Debye for the singlet state
and $D_0$ = 0.5 Debye for the triplet state (where $a_0$ = Bohr radius ).
For numerical illustration, we have chosen PA transitions to the electric field-dressed bound states of excited B$^1\Pi$ potential of LiCs molecule evolving from field-free high ($v'$ = 26) or low ($v'$ = 4) lying vibrational states.
As discussed by Deiglmayr $\it {et.}$ $\it {al.}$ \cite{deiglmayrnjp}, deeply bound excited states (such as $v'$ = 4) of LiCs have pure singlet character and hence spin-orbit coupling can be safely neglected for such states.
Though the bound state with $v'$ = 26 is relatively close to the dissociation threshold and therefore can be influenced by spin-orbit coupling, however the triplet component of this bound state is far less than the singlet one \cite{deiglmayrnjp}.
The potential energy data and permanent dipole moment function of B$^1\Pi$ state are taken from Refs. \cite{grocholajcp2009} and \cite{beriche}, respectively.
We choose lowest hyperfine channel as the single channel approximation \cite{debashree} for our numerical calculation. 
The transition dipole moment is taken to be 4.0 a.u. \cite{deiglmayr} for our numerical calculation.

%==========================================================================================================================================
%Figure-6
\begin{figure}
\includegraphics[width=4.25in]{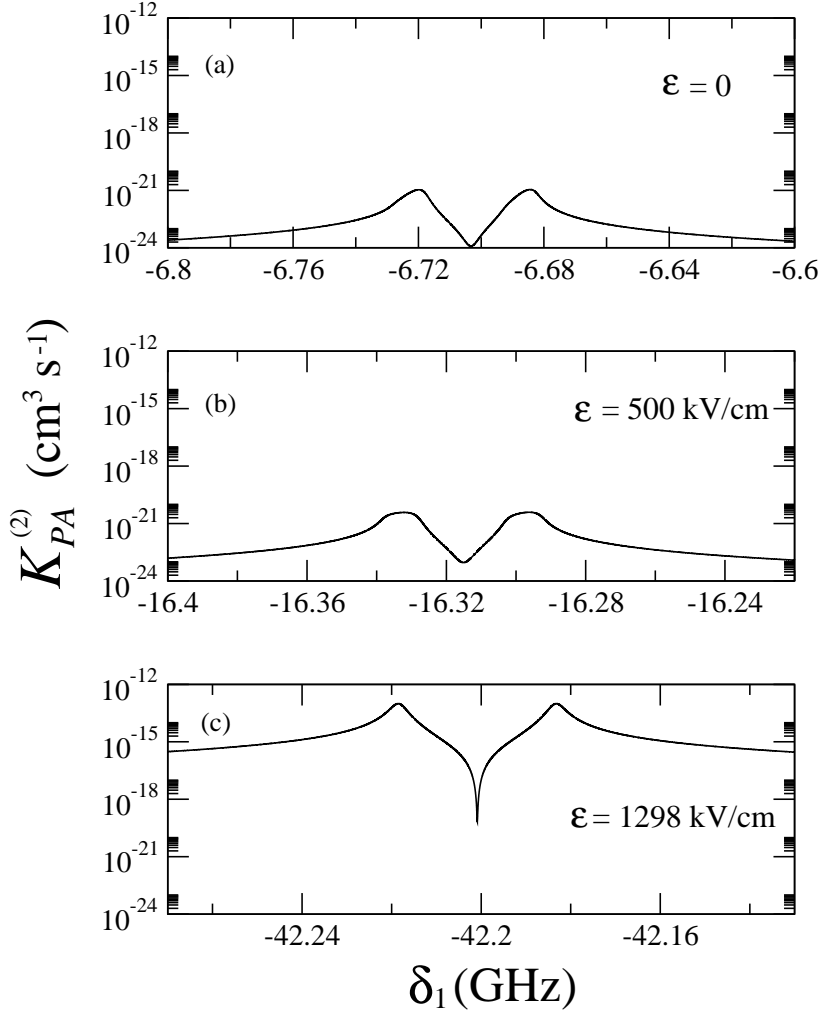}
\caption{Two-color PA rate $K_{PA}^{(2)}$ is plotted as a function of detuning $\delta_1$ (in GHz) of laser-1 at $\cal E$ = 0 (a), 500 kV/cm (b) and $\cal E$ = 1298 kV/cm (c) for transition between states evolving from field-free $J'$=1 of $v'$ = 4 and $J$= 0 of $v$ = 4, keeping laser-2 fixed at resonance.}
\end{figure}
%==========================================================================================================================================

We first calculate the shifts in binding energies of bound states due to electric field effects as schematically shown in figure 1.
The binding energies $E_{v,j,M}$ and $E_{v',j',M'}$ of ground and excited bound states are measured from the threshold of the respective potentials and so both are negative.
Explicitly, shifts are given by
$\Delta'_{v',j',M'}({\cal E}) = -E_{v',j',M'}({\cal E})+E_{v',j',M'}(0)$ 
and $\Delta_{v,j,M}({\cal E}) = -E_{v,j,M}({\cal E})+E_{v,j,M}(0)$ for the excited and ground bound state, respectively.
We tabulate these quantities in Table I for different electric field strengths.
In Table II we display electric field variation of the detuning $\nu_{v,j,M,v',j',M'}$($\cal E$) = $\hbar \nu_A$ - $E_{vjM}(\cal E)$ + $E_{v'j'M'}(\cal E)$ for the transition between states evolving from field-free $v$=28, $j$=0 and $v'$=26, $j'$=1. 
Here $\nu_A$ is the atomic transition frequency between S$_{1/2}$ to P$_{3/2}$ level of Cs.
The change in bound-bound transition frequency with respect to the field-free case $\Delta\nu$($\cal E$) = $\nu_{vjM,v'j'M'}(\cal E)$-$\nu_{vjM,v'j'M'}(0)$ is also presented in Table II.
%========================================================================================================================================
%Figure-7
\begin{figure}
\includegraphics[width=4.25in]{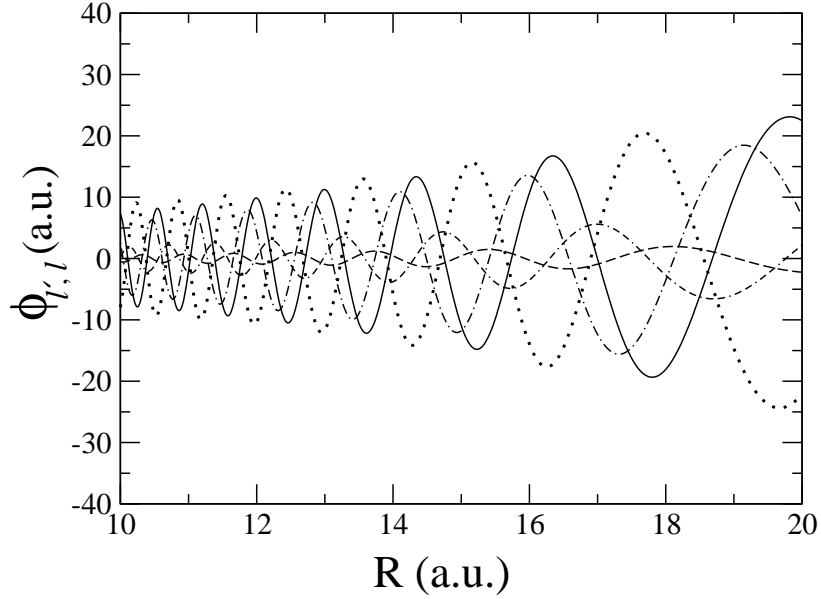}
\caption{Energy normalized scattering wave functions $\phi_{\ell',\ell}$ (in a.u.) is plotted as a function of internuclear separation R (in a.u.) for $\ell$ = 0 and $\ell'$ = 0 (solid line), 1 (dashed line), 2 (dotted line), 3 (dashed-dotted line) and 4 (double dashed-dotted line) at $\cal E$ = 500 kV/cm and energy E = 50$\mu$K.}
\end{figure}
%==========================================================================================================================================
The binding energies of a ground and an excited bound state and the transition frequency between them are schematically shown in figure 1. We now discuss the effects of electric field on two-color PA rate involving high-lying vibrational states in both ground and excited potentials. For the chosen excited state $v'$ = 26 we get the maximum bound-bound Franck-Condon overlap for the ground vibrational state $v$ = 28 in the absence of electric field.  
Hence we have chosen $v$ = 28 for calculating two-color PA. 
Figure 2 shows the variation of two-color PA rate as a function of the detuning $\delta_1$ of laser-1 from atomic resonance while keeping laser-2 on resonance with bound-bound transition.
The two-color spectrum in the absence or presence of electric field shows two-peak structure in accordance with Eq. (16).
From figure 2, we find that the two-color PA rate at $\cal E$ = 500 kV/cm increases by 3 orders of magnitude and the minimum of the two-peak spectral structure is shifted by about 9.61 GHz compared to that at field-free case. 
Furthermore, static electric field is found to significantly modify bound-bound Rabi coupling $\Omega_{12}$. This happens because of electric field-dressing of the bound states. The calculated values of $\Omega_{12}$ are 17.65 MHz at $\cal E$ = 500 kV/cm and 7.98 MHz at $\cal E$ =0.
%=======================================================================================================================================
%Figure-8
\begin{figure}
\includegraphics[width=4.25in]{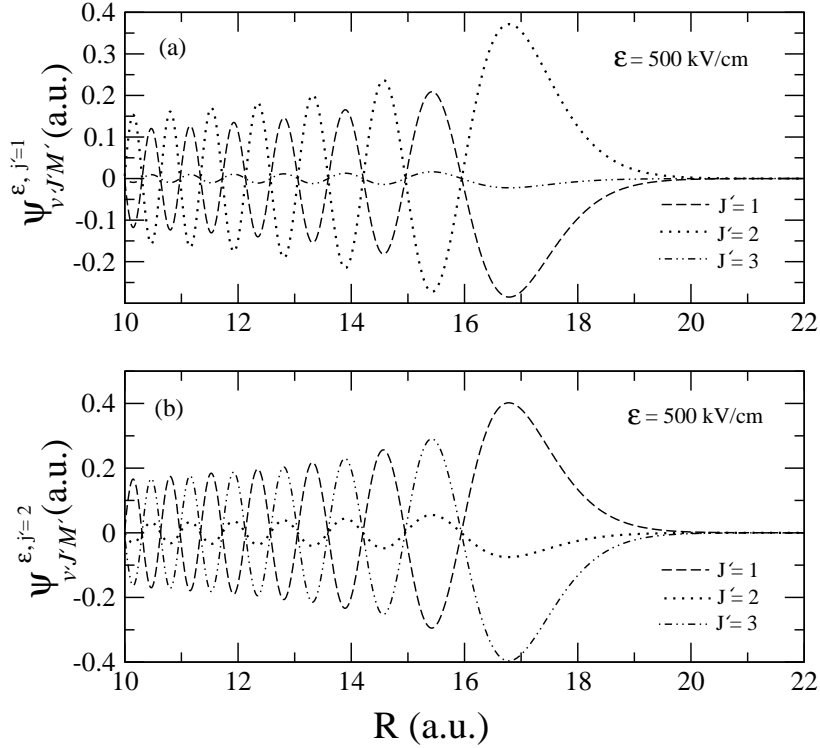}
\caption{Normalized bound state wave functions ${\bf \psi}^{{\cal E},j'}_{v',J',M'}(R)$ (in a.u.) in excited electronic potential evolving from field-free $v'$ = 26, $j'$ = 1 (a) and $j'$ = 2 (b) are plotted as a function of internuclear separation $R$ (in a.u.) for electric field
strength $\cal E$ = 500 kV/cm.}
\end{figure}
%================================================================================================================================
In figure 3 we have plotted the two-color PA rate as a function
of $\delta_2$ = $\hbar\omega_2-(E_{v',j',M'}(0) + \hbar\nu_A - E_{v,j,M}(0))$ for $\cal E$ = 500 kV/cm. This shows a shift of about 16.3 GHz due to static electric field effects. It is worthwhile to mention that the change in the threshold of atom-atom potentials 
by the second order Stark effect at $\cal E$ = 500 kV/cm is of the order of 100 MHz which is much smaller than the shifts of the bound states and bound-bound frequencies for the same electric field. We have therefore neglected this second order Stark effect. 
The static electric field-induced changes in binding energies of the bound states as shown in Table I and II are consistent with the shifts in PA spectrum of figures 2 and 3. The change in binding energies depend on $j'$ or $j$ values, that is, from which field-free rotational state electric-field dressed multichannel bound states have evolved. 
%=============================================================================================================
% Figure-9
\begin{figure}
\includegraphics[width=4.25in]{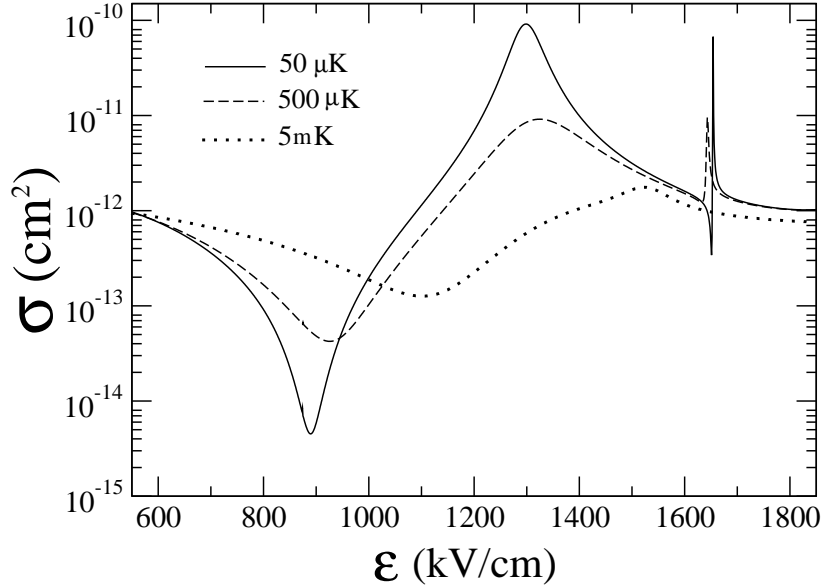}
\caption{Total elastic cross-section $\sigma$ in cm$^2$ vs electric field $\cal E $ in kV/cm is plotted for three different collisional energies E = 50$\mu K$ (solid line), 500 $\mu K$ (dashed line) and 5 mK (dotted line).
}
\end{figure}
%===============================================================================================================
We also plot the one-color PA rate $K_{PA}^{(1)}$ in figure 4 and compare this with the results of figure 2.
One-color PA spectrum at $\cal E$ = 500 kV/cm exhibits similar enhancement and shift as in two-color case. 
This suggests that the enhancement in two-color PA rate is primarily due to electric field effects on collisional states of
ground state continuum. Though we have shown selective results on two-color PA spectrum at $\cal E$ = 500 kV/cm, we have found PA rate enhancement
by 2 or 3 orders of magnitude and significant shifts for electric fields ranging from 100 to 500 kV/cm when optical transitions occur
between high-lying vibrational states in both electronically ground and excited configurations. 
As an example, in figure 5 we have plotted the two-color PA rate as a function of $\delta_1$ at $\cal E$ = 200 kV/cm and $\cal E$ = 500 kV/cm. At $\cal E$ = 200 kV/cm there is also enhancement of two-color PA rate by the same order as in 500
kV/cm, the difference being only a factor of about 2.

We next consider two-color PA involving low-lying vibrational states in both excited and ground molecular potentials.
For this we choose two bound states with $v'$ = 4 and $v$ = 4 for which the two-color PA rate as a function of $\delta_1$ is plotted in figure 6.
Comparing figure 6 with figure 2, we find that for $\cal E$ = 500 kV/cm the enhancement of PA rate involving low-lying vibrational states is small unlike that in case of high-lying vibrational states.
This is because of huge disparity in free-bound Franck-Condon overlap integrals in the two cases. However, at resonant electric field $\cal E$ = 1298 kV/cm the enhancement is quite large even in the case of low-lying vibrational states, as demonstrated in figure 6.

To explain the enhancement of two-color as well as one-color PA rate at lower electric field strengths ($\ll$ 1298 kV/cm), we have plotted in figure 7 different components $\phi_{\ell',\ell}$ of the continuum wave function as given by Eq. (10). 
The different components of a field-dressed bound state in excited potential are displayed in figure 8.  
Due to multichannel nature of the bound state, there are different components of the bound states corresponding to different rotational numbers $J$.  As a result, 
PA spectrum will consist of contributions from different $J$ values.
%==========================================================================================================================================
%Figure-10
\begin{figure}
\includegraphics[width=4.25in]{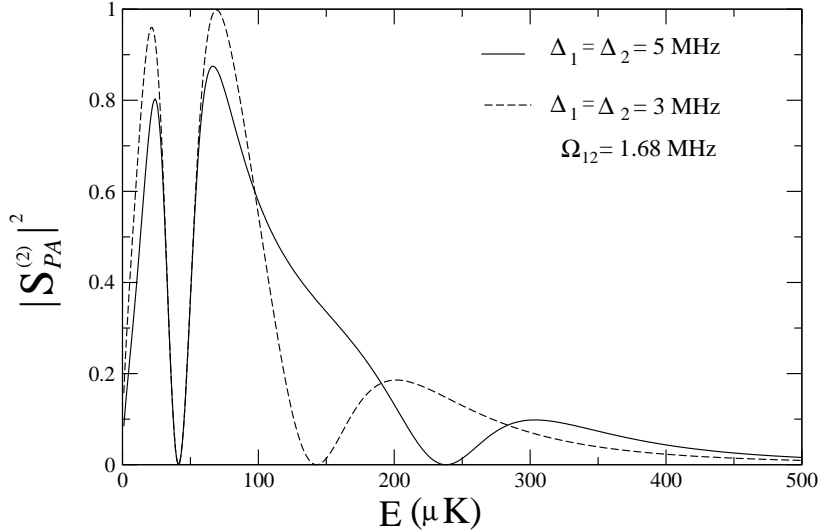}
\caption{$\mid S_\textit{PA}^{(2)}\mid^2$ for two-color PA is plotted as a function of collisional energy E (in $\mu$K) for $\cal E$ = 1298kV/cm. The detunings are set at $\Delta_1$ =  $\Delta_2$ = 5 MHz (solid line) and $\Delta_1$ =  $\Delta_2$ = 3 MHz (dashed line) for $\Omega_{12}$ = 1.68 MHz.}
\end{figure}
%==========================================================================================================================================
To know whether scattering cross section is related to the enhancement in PA rate, we plot the total elastic cross-section $\sigma$ as a function of electric field strength $\cal E$ for 
three different collisional energies in figure 9. 
A prominent anisotropic resonant peak appears for ${\cal E} = {\cal E}_r$ = 1298 kV/cm at low energy. It has been shown that 
this scattering resonance leads to the resonant enhancement in PA rate implying that the resonance is linked to the enhancement of the amplitude of field-dressed multichannel scattering wave function at relatively short separations where PA transitions occur according to Franck-Condon principle.
From scattering theoretical point of view, PA spectrum describes inelastic scattering rate. 
Near an anisotropic resonance inelastic rate can be drastically increased or decreased depending on the superposition of the different outgoing channel states that are strongly coupled to the lossy (PA) channel. 
This is similar to the superposition effects in inelastic rates in magnetic Feshbach resonance as analyzed by Hutson $\it {et.}$ $\it {al.}$ \cite{hutsonprl}.

Finally, we analyze two-color PA probability $\mid S_\textit{PA}^{(2)}\mid^2$ as given in Eq. (15) at resonant electric field $\cal E$ = 1298 kV/cm.
In figure 10 we have plotted the two-color PA probability as a function of collision energy $E$.
Due to large enhancement of $\Gamma$ at low energy, saturation occurs and a three-peak structure is observed in the two-color PA probability rather than its usual two-peak structure. With changing detunings of the lasers the position of the peak which appears at higher energy changes but the peak which appears at lower energy remains the same.
Therefore, the two-peak structure in the lower energy regime appears due to saturation effects and the peak at higher energy is due to the two-color PA process. One of the peaks of two-color PA spectra is merged with the peak which appears due to saturation.        
In figure 11, the detunings are $\Delta_1$ =  $\Delta_2$ = 0.5 MHz (solid line) and $\Delta_1$ =  $\Delta_2$ = 1.5 MHz (dashed line), which correspond to the energy of the first and the second peak of figure 10, respectively. 
%==========================================================================================================================================
%Figure-11
\begin{figure}
\includegraphics[width=4.25in]{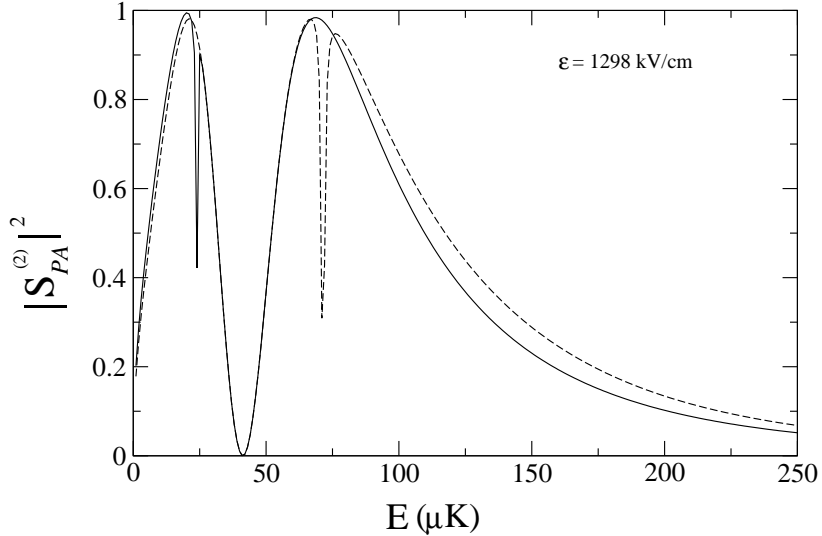}
\caption{$\mid S_\textit{PA}^{(2)}\mid^2$ for two-color PA is plotted as a function of collisional energy E (in $\mu$K) for $\cal E$ = 1298kV/cm. The detunings are set at $\Delta_1$ =  $\Delta_2$ = 0.5 MHz (solid line) and $\Delta_1$ =  $\Delta_2$ = 1.5 MHz (dashed line) which corresponds to the energy of the first and the second peak of figure 10, respectively.}
\end{figure}
%==========================================================================================================================================
\section{Conclusions}
In conclusion, we have presented a non-perturbative method for calculating bound state wave functions in presence of an
external static electric field, for applications in photoassociation between different atoms. In presence of a strong electric field, different rotational states are coupled and therefore a multichannel method is required to solve the problem. 
We have used field-dressed wave functions to calculate  PA spectra.
In the specific case of Li + Cs system both one- and two-color PA rates are enhanced by several orders of magnitude for
electric field strengths of a few hundred kV/cm when PA transitions occur to the high-lying bound states in 
excited electronic potential.
Significant spectral shifts are shown to arise from electric-field dressing of the continuum and bound states. 
We have demonstrated that near electric field induced resonances, a three peak spectrum appears in two-color PA.
Among the two dips of the three-peak spectrum, one is due to saturation effect and the other is due to quantum interference of two transition pathways of two-color PA. 
The sharp feature of the dip of figure 11 is an indication that with the use of static electric field induced anisotropic resonances under appropriate detuning conditions at ultralow temperatures,
ground state polar molecules may be formed in some specific selective ro-vibrational states by two-color PA. 
In this paper, we have carried out our calculations of PA spectrum for low intensity PA lasers. It would be interesting to investigate PA of different atoms in intense laser fields in the presence of a static electric field of
moderate strength of experimental relevance for controlling rotation-vibration coupled molecular states.
Static electric field strength ranging from 100 to 500 kV/cm \cite{2004} are of current interest. There are practical  
difficulties to use an electric field larger than 500 kV/cm in an experiment with cold atoms. 
There is an alternative way to produce nonperturbative electrical effects in scattering states by controlling a shape resonance with a non-resonant intense laser field as proposed by Gonz$\acute{a}$lez-f$\acute{e}$rez and Koch \cite{Gonzalezpra2012}.
To induce an equivalent anisotropic resonance effect between two similar atoms with a laser field, the required intensity of laser is of the order of 10$^{10}$ W cm$^{-2}$ which is quite high.
For two different atoms a laser of moderate intensity ($\sim$ 1 GW cm$^{-2}$) would suffice to induce this type of resonance. Furthermore, in case of two different atoms, there will be combined effects of permanent and induced dipole interactions. 
However, due to dynamical nature of the laser field, the possibility of transitions among different rotational states in ground state manifold can not be ruled out.
Nevertheless, it would be an interesting pursuit to device an alternative approach to realize the discussed effects with a moderately intense laser field of appropriately chosen frequency.

$\bf{Acknowledgment}$

One of us (Debashree Chakraborty) is grateful to Council of Scientific and Industrial Research (CSIR), Government of India, for a support under grant number 09/080(0641)/2009-EMR-I.


\begin{thebibliography}{2}

\bibitem{krems1} R. V. Krems, Phys. Rev. Lett. {\bf 96}, 123202 (2006)

\bibitem{krems2} Z. Li and R.V. Krems, Phys. Rev. A. {\bf 75}, 032709 (2007)

\bibitem{debashree} D. Chakraborty, J. Hazra and B. Deb, J. Phys. B.: At. Mol. Opt. Phys. {\bf 44}, 095201 (2011)

\bibitem{bigelow} C. Haimberger, J. Kleinert, M. Bhattacharya and N. P. Bigelow, Phys. Rev. A. {\bf 70}, 021402(R)(2004)

\bibitem{marcassa} M. W. Mancini, G. D. Telles, A. R. L. Caires, V. S. Bagnato and L. G. Marcassa, Phys. Rev. Lett. {\bf 92}, 133203 (2004)

\bibitem{k.k.ni} K.-K. Ni, S. Ospelkaus, M. H. G. de Miranda, A. Pe$'$er, B. Neyenhuis, J. J. Zirbel, S. Kotochigova, P. S. Julienne, D. S. Jin and J. Ye, Science {\bf 322}, 231 (2008)

\bibitem{kerman} A. J. Kerman, J. M. Sage, S. Sainis, T. Bergeman and D. DeMille, Phys. Rev. Lett. {\bf 92}, 153001 (2004)

\bibitem{sage} J. M. Sage, S. Sainis, T. Bergeman and D. DeMille, Phys. Rev. Lett. {\bf 94}, 203001 (2005)

\bibitem{kraft} S. D. Kraft, P. Staanum, J. Lange, L. Vogel, R. Wester and M. Weidem$\ddot{u}$ller, J. Phys. B.: At. Mol. Opt. Phys. {\bf 39}, S993 (2006)

\bibitem{deiglmayr} J. Deiglmayr, A. Grochola, M. Repp, K. M$\ddot{o}$rtlbauer, C. Gl$\ddot{u}$ck, J. Lange, O. Dulieu, R. Wester and M. Weidem$\ddot{u}$ller, Phys. Rev. Lett {\bf 101}, 133004 (2008)

\bibitem{wedimullerchemrev2012} J. Ulmanis, J. Deiglmayr, M. Repp, R. Wester, and M. Weidem$\ddot{u}$ller, Chem. Rev. {\bf 112}, 4890 (2012).
 
\bibitem{takekoshipra2012} T. Takekoshi, M. Debatin, R. Rameshan, F. Ferlaino, R. Grimm, H. C. N$\ddot{a}$gerl, C. Ruth Le Sueur, J. M. Hutson, P. S. Julienne, S. Kotochigova, and E. Tiemann, Phys. Rev. A. {\bf 85}, 032506 (2012).

\bibitem{shapirochemrev2012} C. P. Koch and M. Shapiro, Chem. Rev. {\bf 112}, 4928 (2012).

\bibitem{Gonzalezpra2012} R. Gonz$\acute{a}$lez-F$\acute{e}$rez and C. P. Koch, Phys. Rev. A. {\bf 86}, 063420 (2012)

\bibitem{meyenn} K. von Meyenn, Z. Phys. {\bf 231}, 154 (1970)

\bibitem{2004} R. Gonz$\acute{a}$lez-F$\acute{e}$rez and P. Schmelcher, Phys. Rev. A {\bf 69}, 023402 (2004)

\bibitem{2005} R. Gonz$\acute{a}$lez-F$\acute{e}$rez and P. Schmelcher, Phys. Rev. A {\bf 71}, 033416 (2005)

\bibitem{gonzalezpra2007} R. Gonz$\acute{a}$lez-F$\acute{e}$rez, M. Weidem$\ddot{u}$ller and P. Schmelcher, Phys. Rev. A {\bf 76}, 023402 (2007)

\bibitem{gonzaleznjp2009} R. Gonz$\acute{a}$lez-F$\acute{e}$rez, and P. Schmelcher, New. J. phys. {\bf 11}, 055013 (2009)

\bibitem{gonzalezpccp2011} R. Gonz$\acute{a}$lez-F$\acute{e}$rez, and P. Schmelcher, Phys. Chem. Chem. Phys. {\bf 13}, 18810 (2011)

\bibitem{johnsonjcp1978} B. R. Johnson, J. Chem. Phys. {\bf 69}, 4678 (1978)

\bibitem{supplemental} See supplementary material at [] for the method of calculation of multichannel bound state. 

\bibitem{Gordon} R. G. Gordon, J. Chem. Phys. {\bf 51}, 14 (1969)

\bibitem{napolitano} R. Napolitano, J. Weiner, C. J. Williams and P. S. Julienne, Phys. Rev. Lett. {\bf 73}, 1352 (1994)

\bibitem{julienne} J. L. Bohn and P. S. Julienne, Phys. Rev. A {\bf 54}, R4637 (1996)

\bibitem{stannumpra2007} P. Stunnum, A. Pashov, H. Knockel and E. Tiemann, Phys. Rev. A {\bf 75}, 042513 (2007)

\bibitem{aymer} M. Aymar and O. Dulieu, J. Chem. Phys. {\bf 122}, 204302 (2005)

\bibitem{deiglmayrnjp} J. Deiglmayr, P. Pellegrini, A. Grochola, M. Repp, R C$\hat{o}$t$\acute{e}$, O. Dulieu, R. Wester and M. Weidem$\ddot{u}$ller, New. J. Phys. {\bf 11}, 055034 (2009) 

\bibitem{grocholajcp2009} A. Grochola, A. Pashov, J. Deiglmayr, M. Repp, E Tiemann and R. Wester, J. Chem. Phys. {\bf 131} 054304 (2009)

\bibitem{beriche} N. Mabrouk, H. Berriche, H. Ben Ouada and F. X. Gadea, J. Phys. Chem. A {\bf 114} 6657 (2010)

\bibitem{hutsonprl} J. M. Hutson, M. Beyene and M. L. Gonz$\acute{a}$lez-Mart$\acute{i}$nez, Phys. Rev. Lett. {\bf 103}, 163201 (2009)

\end{thebibliography}
\end{document}